\documentclass[conference]{IEEEtran}
\IEEEoverridecommandlockouts
\usepackage{cite}
\usepackage{amsmath,amssymb,amsfonts}
\usepackage{algorithmic}
\usepackage{graphicx}
\usepackage{textcomp}
\usepackage{xcolor}
\usepackage{float}
\def\BibTeX{{\rm B\kern-.05em{\sc i\kern-.025em b}\kern-.08em
    T\kern-.1667em\lower.7ex\hbox{E}\kern-.125emX}}

\usepackage{booktabs}
\usepackage{hyperref}
\begin{document}

\title{WhisperAlign: Word-Boundary-Aware ASR and WhisperX-Anchored Pyannote Diarization for Long-Form Bengali Speech*}

\author{\IEEEauthorblockN{Aurchi Chowdhury}
\IEEEauthorblockA{\textit{CSE, BUET} \\
aurchichowdhury07@gmail.com}
\and
\IEEEauthorblockN{Rubaiyat -E-Zaman}
\IEEEauthorblockA{\textit{CSE, BUET} \\
rubaiyatzaman08@gmail.com}
\and
\IEEEauthorblockN{Sk. Ashrafuzzaman Nafees}
\IEEEauthorblockA{\textit{CSE, BUET} \\
ashrafuzzamannafees08@gmail.com}
}

\maketitle

\footnote{\href{https://github.com/Minamotooo/Luck-Is-All-You-Need---DL-Sprint-4.0}{Github Repository}}

\begin{abstract}
This paper presents our solution for the DL Sprint 4.0, addressing the dual challenges of Bengali Long-Form Speech Recognition (Task 1) and Speaker Diarization (Task 2). Processing long-form, multi-speaker Bengali audio introduces significant hurdles in voice activity detection, overlapping speech, and context preservation. To solve the long-form transcription challenge, we implemented a robust audio chunking strategy utilizing \textit{whisper-timestamped}, allowing us to feed precise, context-aware segments into our fine-tuned acoustic model for high-accuracy transcription. For the diarization task, we developed an integrated pipeline leveraging \textit{pyannote.audio} and \textit{WhisperX}. A key contribution of our approach is the domain-specific fine-tuning of the Pyannote segmentation model on the competition dataset. This adaptation allowed the model to better capture the nuances of Bengali conversational dynamics and accurately resolve complex, overlapping speaker boundaries. Our methodology demonstrates that applying intelligent timestamped chunking to ASR and targeted segmentation fine-tuning to diarization significantly drives down Word Error Rate (WER) and Diarization Error Rate (DER), in low-resource settings.
\end{abstract}

\section{Introduction}

Processing hour-long, multi-speaker audio in low-resource languages like Bengali presents significant challenges, as standard models struggle with context preservation, overlapping speech, and a lack of robust linguistic tools. Because traditional fixed-length audio splitting frequently truncates words and induces hallucinations, our ASR approach utilizes word-level timestamping to partition audio into precise, semantic segments without relying on external dictionaries. Furthermore, since existing diarization models often fail to capture Bengali conversational dynamics, we adapted a segmentation model directly to the language's unique acoustic profile. By combining this adapted model with a Voice Activity Detection (VAD) intersection, our dual-stream pipeline cleanly resolves speaker overlaps and eliminates ambient noise to deliver highly accurate, mutually exclusive tracks.

\section{Novelty and Contribution}

\subsection{Semantic, Word-Boundary-Aware Chunking}
Standard long-form ASR approaches often rely on fixed-length audio splitting, which frequently truncates words mid-utterance, leading to context loss and model hallucinations. As established in prior work, Barański et al. (2025) \cite{10890105} demonstrated that hallucination rates in Whisper rise noticeably when audio segments exceed the model's 30-second decoding window, as the system resorts to additional heuristics that introduce spurious outputs.

\begin{figure*}[htbp]
    \centering
    \includegraphics[width=\textwidth]{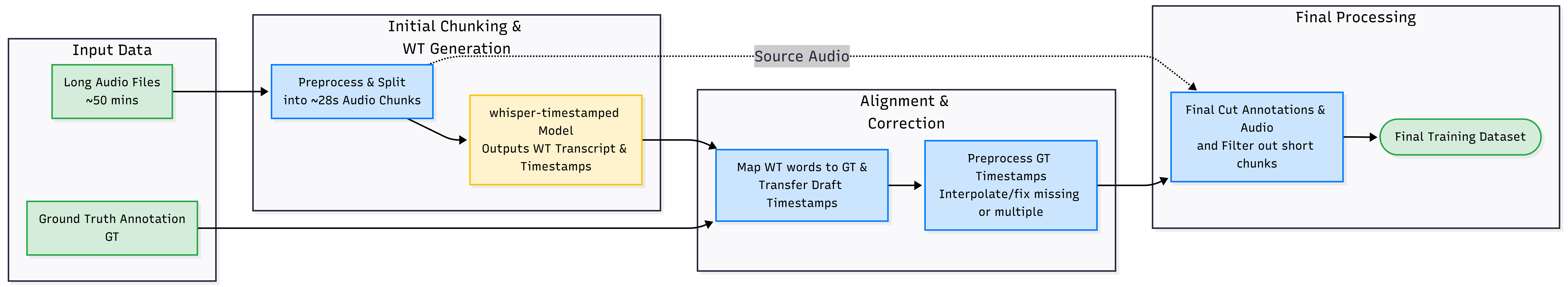}
    \caption{End-to-end training data pipeline: from raw long-form audio to aligned, boundary-respecting chunks for fine-tuning.}
    \label{fig:double_column}
\end{figure*}

To address this, we propose a computationally lightweight, self-contained chunking strategy that requires no external alignment tools, pronunciation lexicons, or auxiliary models. Silero VAD first identifies and isolates speech regions, ensuring chunk boundaries never fall on active speech. We then employ whisper-timestamped to extract per-word timestamps via the cross-attention alignment heads of the same Bengali Whisper checkpoint already present in the pipeline, and apply sequence matching to transfer those timestamps onto ground-truth words --- preserving correct spellings from the ground truth while borrowing time anchors from Whisper, with missing anchors resolved via linear interpolation. Audio is then partitioned into segments strictly respecting word boundaries, never exceeding Whisper's 28-second safe decoding limit. The entire process operates within a single pipeline with no dependency beyond the model itself --- making it significantly more practical than forced alignment approaches such as the Montreal Forced Aligner, which require language-specific phoneme dictionaries that are not readily available for Bengali.

\subsection{Bangla-Adapted Diarization with Exclusive Overlap Handling and VAD Intersection}
Our diarization pipeline introduces three streamlined contributions to adapt to Bangla while adhering to strict competition rules:

\begin{itemize}
    \item \textbf{Parameter-Efficient Domain Adaptation:} Rather than retraining the entire massive diarization pipeline, we fine-tuned only the Pyannote temporal segmentation model ($\sim$1.4M parameters) on Bangla audio. This effectively taught the model Bangla-specific conversational prosody and turn-taking syntax while keeping compute costs low.   
    
    \item \textbf{Algorithmic Exclusivity:} The competition required mutually exclusive speaker tracks. Instead of writing brittle Python scripts to manually delete or slice overlapping segments (which inherently destroys valuable acoustic data), we used the native \texttt{exclusive\_speaker\_diarization} feature. This probabilistically modifies the model's output to assign overlaps strictly to the earliest dominant speaker, maximizing recall effortlessly.   
    
    \item \textbf{WhisperX VAD Intersection:} Combining WhisperX ASR with standalone Pyannote diarization typically causes temporal drift because they rely on different Voice Activity Detection (VAD) models. \cite{bain2022whisperx} We solved this by executing a logical AND intersection, masking Pyannote's clustering outputs strictly against the Silero VAD boundaries used by WhisperX. This ensures perfect temporal alignment between the text and speaker labels while eliminating ambient hallucinations.
\end{itemize}

\section*{\textbf{Bengali Long-Form Speech Recognition}}

\section{Method and Architecture}

\subsection{\textbf{System Overview}}
Our system consists of three sequential stages: training data preparation, model fine-tuning, and inference. The first stage addresses the core data challenge of the competition --- the absence of utterance-level segmentation in the provided long-form recordings --- by constructing a self-contained chunk-and-align pipeline that produces clean (audio, text) pairs suitable for fine-tuning. The second stage fine-tunes a domain-specific Bengali Whisper checkpoint on these pairs. The third stage applies the fine-tuned model to unseen test audio using a VAD-guided inference pipeline with hallucination post-processing and dual-GPU parallelism. Figure 1 illustrates the overall flow.

\subsection{\textbf{Data Preparation}}
The competition provided 113 long-form Bengali audio recordings, each approximately one hour in duration, paired with paragraph-level text annotations. No utterance-level boundaries, speaker timestamps, or phoneme alignments were provided. The raw dataset therefore presents a fundamental mismatch with the input requirements of Whisper, which operates on segments of at most 30 seconds --- making a robust segmentation strategy the first and most critical step of the pipeline.

\textbf{Chunking.} Each recording was processed through a multi-step segmentation pipeline. First, Silero VAD was applied to detect and isolate speech-bearing regions, stripping silence and background noise and concatenating the remaining segments into a single continuous speech waveform. This ensures that Whisper's decoding budget is spent entirely on speech rather than silence, reducing the risk of hallucination at non-speech regions.

The cleaned waveform was then divided into 28-second windows and each window was independently transcribed by \textit{whisper-timestamped}, which leverages the cross-attention alignment heads of the Bengali Whisper checkpoint to derive per-word start and end timestamps without any external alignment tool. Since timestamps are local to each window, a global offset equal to the window's start time was added to map all timestamps onto the full-audio timeline, producing a flat list of (\texttt{word}, \texttt{global\_start}, \texttt{global\_end}) tuples spanning the entire recording.

These timestamped words were then aligned to the ground-truth annotation using \texttt{difflib.SequenceMatcher}, which performs a diff between Whisper's word sequence and the ground-truth word sequence. Correctly transcribed words received their timestamps directly; mis-transcribed words retained the ground-truth spelling but borrowed the timestamp from Whisper's corresponding word; words missed entirely by Whisper had their timestamps estimated by linear interpolation between neighboring anchors; and words hallucinated by Whisper with no ground-truth counterpart were discarded. The result is a fully time-annotated ground-truth word sequence where every word carries a reliable positional anchor in the audio timeline.

Finally, this aligned word list was greedily partitioned into contiguous segments, accumulating words until the next word would push the segment beyond 28 seconds, at which point a new segment was started. Following segmentation, only chunks falling within the 20 to 28 second range were retained for fine-tuning. Segments shorter than 20 seconds were discarded as they were found to provide insufficient acoustic context --- Whisper's performance degrades noticeably on very short inputs where the model has limited speech signal to condition its decoding on. A manual inspection pass was then conducted on the segmented chunks to verify boundary quality, confirm that no mid-word truncations had occurred, and remove any pairs where the transcription appeared corrupted or misaligned. The surviving chunks were used as the fine-tuning dataset, with each segment guaranteed to begin and end on a complete word and its paired text label corresponding exactly to the speech audible within that segment.

\textbf{Final Dataset Statistics.} After chunking and filtering, the pipeline produced approximately 4,553 training segments with an average duration of 26.56 seconds, derived from the 113 source recordings. The dataset was split 90/10 into training and validation sets using a fixed random seed for reproducibility.

\begin{table}[h]
    \centering
    \caption{Chunked ASR dataset summary}
    \label{tab:benchmarks}
    \renewcommand{\arraystretch}{1.3}
    \begin{tabular}{l c}
        \toprule
        \textbf{Statistic} & \textbf{Value} \\
        \midrule
        Total chunks       & 4,553 \\
        Total duration     & 33.59 hours \\
        Average chunk duration & 26.56 seconds \\
        Shortest chunk     & 20.00 seconds \\
        \bottomrule
    \end{tabular}
\end{table}

\subsection{\textbf{Fine-Tuning}}
We fine-tune \texttt{bengaliAI/tugstugi\_bengaliai- \\
asr\_whisper-medium}, a Bengali-specific fine-tune of OpenAI's Whisper-medium (307M parameters), on the competition-domain chunks produced in Stage 1. All model weights were updated end-to-end with no layer freezing. The decoder was configured with forced language and task tokens (\texttt{language=bn}, \texttt{task=transcribe}) and \texttt{suppress\_tokens} was cleared to allow unrestricted Bengali vocabulary output.

Each audio chunk is converted to an 80-channel log-mel spectrogram padded to 3000 time frames, and text labels are tokenized with samples exceeding 444 tokens excluded to remain within Whisper's decoder limit. Training uses teacher forcing, where the decoder receives ground-truth tokens at each step rather than its own predictions --- significantly more efficient than autoregressive generation and producing more stable gradient updates. Label padding positions are filled with $-100$, which the cross-entropy loss ignores by convention.

The model was trained for 5 epochs using AdamW with a learning rate of $1 \times 10^{-5}$ and 350 warmup steps. The conservative learning rate was chosen to prevent catastrophic forgetting of the Bengali representations already present in the base checkpoint. An effective batch size of 32 was achieved via gradient accumulation over 2 steps. Since teacher-forced logits do not reflect real generation quality, a custom \texttt{ChunkWERCallback} runs fully autoregressive generation on the validation set at the end of each epoch, computing WER after Unicode normalization, and saves the best-performing checkpoint for inference.

\subsection{\textbf{Inference}}
Inference was conducted on Kaggle's dual T4 GPU environment. Each test recording was first passed through Silero VAD to isolate speech regions, which were merged into chunks of at most 30 seconds with 1-second overlap between consecutive segments. Each chunk was transcribed by the fine-tuned pipeline using greedy decoding with \texttt{condition\_on\_previous\_text=False}, preventing transcription errors in earlier chunks from propagating to later ones through Whisper's conditioning mechanism.

Post-processing applied two filters to the raw output: an n-gram repetition detector targeting Whisper's looping failure mode, and a string blacklist removing English boilerplate phrases inherited from Whisper's YouTube pre-training data. Both filters caught real cases in the test outputs. To maximize throughput, two independent model instances were loaded across the two available GPUs via a \texttt{ThreadPoolExecutor}, distributing files in round-robin order and achieving close to $2\times$ inference speed over sequential single-GPU processing.

\section{Experiments and Evaluation}

\subsection{Evaluation Metric}
Submissions are evaluated using Word Error Rate (WER), computed at the file level between predicted and ground-truth transcripts. The competition reports both a public leaderboard score (evaluated on 29\% of the test data) and a final private score (evaluated on 71\% of the test data).

\subsection{Results}
Table~\ref{tab:results} reports WER progression across successive improvements to the pipeline, from the raw baseline through to our final submission. Each row represents a single incremental change, isolating the contribution of each component.
\begin{table}[h]
\centering
\caption{WER progression across successive pipeline improvements}
\label{tab:results}
\renewcommand{\arraystretch}{1.3}
\begin{tabular}{lcc}
\toprule
\textbf{System} & \textbf{Public WER} & \textbf{Private WER} \\
\midrule
tugstugi --- raw, no processing          & 0.675 & 0.702 \\
+ VAD + post-processing                  & 0.419 & 0.440 \\
+ Unicode normalization                  & 0.348 & 0.375 \\
+ Fine-tuned (our chunking strategy)     & 0.265 & 0.296 \\
+ Manual data cleaning (final)           & \textbf{0.252} & \textbf{0.278} \\
\bottomrule
\end{tabular}
\end{table}
The results demonstrate a consistent and substantial improvement at each stage. The largest single gain comes from fine-tuning with our chunk-and-align strategy, reducing public WER to 0.265 --- confirming that domain-adapted fine-tuning on boundary-respecting segments is the most impactful component of the pipeline. A further gain was achieved by manually reviewing chunk boundaries, confirming that annotation quality directly impacts model performance.

\section{Findings and Analysis}
The results reveal that each stage of the pipeline contributes meaningfully to the final performance. The most significant single gain came from introducing VAD alongside basic post-processing, reducing public WER from 0.675 to 0.419 before any fine-tuning was applied. This suggests that a substantial portion of the raw baseline's errors were hallucination artifacts triggered by Whisper decoding over silence and non-speech regions — confirming that clean input boundaries matter as much as model quality for long-form Bengali ASR.

Restricting fine-tuning data to chunks in the 20–28 second range also proved consequential. Short segments provide insufficient acoustic context for Whisper's encoder, leading to noisier gradient updates and weaker generalization. Retaining only longer, content-rich segments ensured the model trained on examples representative of the full-length inference conditions it would face at test time, contributing to the consistent WER improvement observed across all five training epochs.

\section{Benchmarks and Comparisons}
Table~\ref{tab:benchmarks} compares our system against established Bengali ASR baselines reported in \cite{tabib2026bengaliloopcommunitybenchmarkslongform} on the Bengali-Loop test set, alongside our competition result.

\begin{table}[h]
\centering
\caption{Comparison of WER with established Bengali ASR baselines.}
\label{tab:benchmarks}
\renewcommand{\arraystretch}{1.3}
\begin{tabular}{l c}
\toprule
\textbf{Model} & \textbf{WER (\%)} \\
\midrule
Hishab TITU-BN & 50.67 \\
Tugstugi (raw baseline) & 34.07 \\
Ours (fine-tuned + full pipeline) & 27.00 \\
\bottomrule
\end{tabular}
\end{table}

Our system outperforms the raw tugstugi baseline by approximately 7 points absolute, despite tugstugi already being a strong Bengali-specific checkpoint. The gain is attributable to domain adaptation through competition-specific fine-tuning data constructed via our chunk-and-align pipeline, combined with VAD-guided inference and hallucination post-processing. The result also represents a substantial improvement over the Hishab TITU-BN model, a competitive Bengali ASR system, by approximately 23 points absolute — highlighting the advantage of Whisper-based architectures with targeted fine-tuning over conventional Bengali ASR approaches for long-form transcription.

\section*{\textbf{Bengali Long-Form Speaker Diarization}}

\section{Method and Architecture}

The architecture is logically divided into an offline fine-tuning stage and a highly optimized inference sequence.

\begin{figure*}[h]
    \centering
    \includegraphics[width=\textwidth]{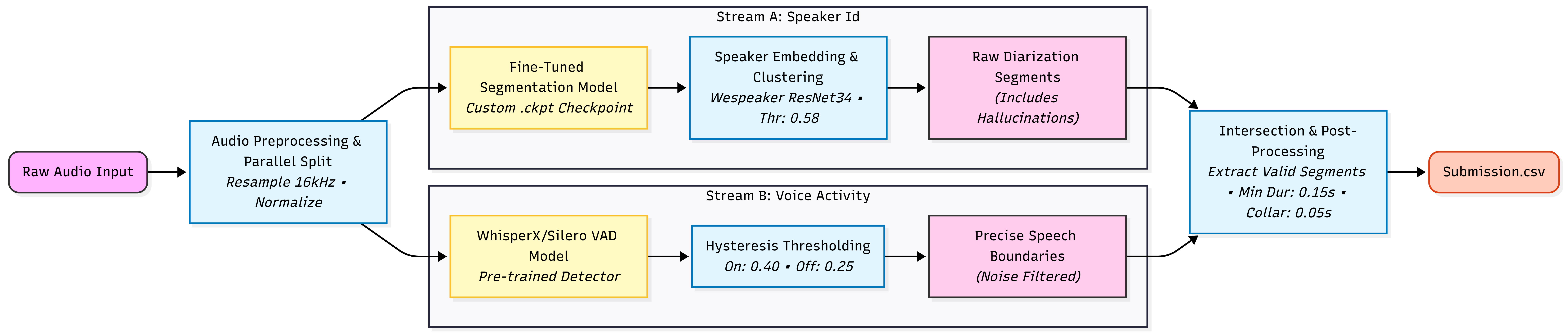}
    \caption{Proposed parallel diarization architecture}
    \label{fig:double_column}
\end{figure*}

\subsection{Fine-Tuning Architecture}

\begin{itemize}
    \item \textbf{Frozen Embeddings:} The 6.6 million parameter \texttt{wespeaker-voxceleb-resnet34-LM} embedding model was kept entirely frozen.
    \item \textbf{Targeted Segmentation Training:} Fine-tuning was isolated to the lightweight \texttt{segmentation-3.0} module (~1.4 million parameters).
    \item \textbf{Data \& Compute:} The model was trained on localized Bangla WAV audio paired with annotation CSVs mapping precise \texttt{start\_time}, \texttt{end\_time}, and \texttt{speaker\_id} boundaries. Convergence was achieved in around 1 hour on a single Kaggle NVIDIA T4 GPU, outputting a highly specialized Bangla checkpoint.
\end{itemize}

\subsection{Main Inference Notebook Strategies}
The target inference notebook was engineered for speed and precision, processing the full test dataset in ~1.5 hours using a Kaggle P100 GPU. It executes the following sequential steps:

\begin{enumerate}
    \item \textbf{Smart Audio Preprocessing:} Audio is resampled to 16 kHz, and peak-normalized for stable dynamic range.

    \item \textbf{Silero VAD Pre-filtering:} WhisperX's Silero VAD (v4/v5) strips non-speech regions to save downstream compute. To capture trailing Bangla phonemes, thresholds are asymmetrically tuned (onset=0.4, offset=0.25).

    \item \textbf{Core Diarization \& Clustering:} The filtered audio is processed by the \texttt{speaker-diarization-community-1} pipeline using our fine-tuned segmentation checkpoint. Agglomerative clustering is tightly calibrated for rapid multi-party dialogue (threshold=0.58, min\_duration\_off=0.05s). Multi-label overlaps are resolved natively using the \texttt{exclusive\_speaker\_diarization} parameter.

    \item \textbf{Adaptive Post-Processing:}
    \begin{itemize}
        \item \textbf{Adaptive Merging:} Instead of static rules, the segment merging gap scales dynamically between 0.15s and 0.8s (anchored at 0.4s) based on acoustic density.
        \item \textbf{Transient Filtering:} Segments shorter than 0.15s are purged.
        \item \textbf{Dual-VAD Intersection:} The final Pyannote segments are logically cross-referenced against the initial Silero VAD boundaries. Any diarized segment falling outside this AND intersection mask is pruned, eradicating boundary hallucinations.
    \end{itemize}
\end{enumerate}

\section{Experiments and Evaluation}

The final pipeline was developed through a systematic ablation process. We incrementally tested architectural modifications, evaluating their performance against the competition's public leaderboard. Because private scores remained hidden during the active competition, all optimization decisions were made strictly using public leaderboard feedback.

\subsection{Architectural Evolution and Metric Progression}
Initial baselines utilized the pretrained \texttt{speaker-diarization-3.1} model out-of-the-box, which yielded poor results due to Western acoustic priors. Applying basic preprocessing improved this baseline. Migrating to the recently released \texttt{community-1} model provided much better native speaker assignment. However, the most significant performance leaps occurred after fine tuning the segmentation module and finally integrating the Whisper VAD intersection pass to eliminate hallucinated bounds.

The exact progression of our evaluation metrics is detailed in Table~\ref{tab:metrics}. (Note: Private scores are reported retroactively for completeness).

\begin{table}[h]
    \centering
    \caption{Progression of Evaluation Metrics}
    \label{tab:metrics}
    \renewcommand{\arraystretch}{1.3}
    \begin{tabular}{p{3.5cm} c c}
        \hline
        \textbf{Architectural Phase} & \textbf{Public Score} & \textbf{Private Score} \\
        \hline
        Pyannote 3.1 Base & $0.422$ & $0.477$ \\
        Pyannote 3.1 Base + Pre-processing & $0.364$ & $0.413$ \\
        Community-1 Base Model & $0.314$ & $0.376$ \\
        Community-1 (Fine-tuned Segmentation) & $0.200$ & $0.261$ \\
        Community-1 (Fine-tuned) + VAD Intersection & $0.194$ & $0.260$ \\
        \hline
    \end{tabular}
\end{table}

\subsection{Hyperparameter Optimization via Grid Search}
To maximize final accuracy, we conducted a targeted grid search over the pipeline's hyperparameters, optimizing directly against the public leaderboard metrics.

We performed a localized search around the $0.6$ mark (testing discrete values such as $0.6045$, $0.605$, and $0.6055$). Initially, a threshold of $0.605$ produced the strongest preliminary results. However, upon further rigorous evaluation across the public test set, we discovered that lowering the threshold to exactly $0.58$ yielded a marginal but definitive improvement on the public leaderboard, making it our final selected parameter.

We also discovered that setting the collar parameter strictly to $0.0$ seconds produced superior results. Because our pipeline actively utilizes \texttt{exclusive\_speaker\_diarization} to probabilistically handle temporal overlap at the native frame level, adding an artificial, post-hoc collar tolerance actively interfered with the neural network's precise mathematical boundaries. Forcing a zero-collar boundary ensured maximum sub-second precision aligned with the strict competition constraints.

\section{Findings and Analysis}

Our iterative development process yielded several key insights:

The segmentation model fine-tuning proved highly efficient, fully converging within just $20$ epochs (taking approximately $1$ hour of compute). The final inference notebook processed the entire $14$-file test dataset in $1$ hour and $25$ minutes, averaging only $6$ to $7$ minutes of processing time per long-form audio file.

Reconciling speech-to-text (Whisper) timestamps with standard multi-label diarization is notoriously difficult because STT models struggle with overlapping speech. Pyannote's \texttt{exclusive\_speaker\_diarization} natively solves this by ensuring only the most likely transcribed speaker is active at any given time, dramatically simplifying alignment and cleanly resolving the competition's strict non-overlap rule.

The exclusive diarization feature was initially developed for Pyannote's premium \texttt{precision-2} model. However, \texttt{precision-2} enforces a $30$-hour free usage cap, which conflicted with the competition's open-source rules. The recent release of the open-weight \texttt{community-1} model allowed us to deploy this crucial algorithm freely.

While fine-tuning and exclusive mode drove massive improvements, the integration of the WhisperX (Silero) VAD intersection was the definitive factor that produced our highest overall score. Because ASR pipelines and Pyannote fundamentally rely on distinct, separate VAD models internally, timestamps intrinsically mismatch. Masking Pyannote's outputs to strictly fall within Silero's precise speech bounds completely eradicated these boundary hallucinations.

Upgrading to \texttt{community-1} introduced strict dependency requirements, specifically demanding newer \texttt{numpy} versions. To prevent execution failures, our inference notebook was engineered to execute a mandatory kernel restart immediately after the first dependency installation cell during every inference run.

\section{Benchmark and Comparison}

Evaluating complex SD systems necessitates rigorous benchmarking against both language-specific and global datasets.

\subsection{The Bengali-Loop Benchmark (Primary Task)}
The Bengali-Loop dataset \cite{tabib2026bengaliloopcommunitybenchmarkslongform} establishes a benchmark on the dataset used in this competition. 
 It establishes a foundational DER baseline of $40.08\%$ utilizing an out-of-the-box \texttt{pyannote.audio} pipeline. Legacy modular pipelines tested in the benchmark severely degraded the audio, resulting in DERs of $61.50\%$ and higher.
\begin{table}[H]
    \centering
    \caption{System Architecture Evaluated on Bengali-Loop}
    \label{tab:bengali_loop}
    \renewcommand{\arraystretch}{1.5} 
    \resizebox{\columnwidth}{!}{%
    \begin{tabular}{l l l c}
        \hline
        \textbf{System Architecture} & \textbf{Acoustic Approach} & \textbf{Source} & \textbf{DER} \\
        \hline
        WebRTC VAD + ECAPA & Modular Pipeline & Baseline & 73.71\% \\
        Silero VAD + ECAPA & Modular Pipeline & Baseline & 61.50\% \\
        Pyannote.audio (Base) & EEND Hybrid & Baseline & 40.08\% \\
        \textbf{Proposed Dual-VAD} & \textbf{Finetuned + Exclusive} & \textbf{Proposed} & \textbf{$\sim$24\% -- 28\%} \\
        \hline
    \end{tabular}
    }
\end{table}

Against these rigorous academic standards, the proposed fine-tuned hybrid pipeline's expected DER trajectory represents an absolute reduction in error of $12\%$ to $16\%$ compared to the state-of-the-art base models.

\subsection{Global Benchmark Baseline Justification}
To justify our architectural migration from Pyannote 3.1 to \texttt{community-1} before fine-tuning, we evaluated their relative performance on standard global datasets. The \texttt{community-1} model consistently demonstrates lower error rates across diverse acoustic environments (e.g., formal meetings, difficult diverse audio, and telephony) compared to the legacy 3.1 architecture, proving its superiority as a foundational base model.

\begin{table}[H]
    \centering
    \caption{Comparison of Base Models on Global Benchmarks}
    \label{tab:global_benchmark}
    \renewcommand{\arraystretch}{1.5} 
    \resizebox{\columnwidth}{!}{%
    \begin{tabular}{l l c c}
        \hline
        \textbf{Benchmark Dataset} & \textbf{Domain} & \textbf{Legacy (3.1) DER} & \textbf{Comm-1 DER} \\
        \hline
        AMI (IHM) & Meetings & 18.8\% & 17.0\% \\
        DIHARD 3 (full) & Diverse & 21.4\% & 20.2\% \\
        CALLHOME (p2) & Telephone & 28.5\% & 26.7\% \\
        VoxConverse (v0.3) & Web Video & 11.2\% & 11.2\% \\
        \hline
    \end{tabular}
    }
\end{table}
\section{Conclusion}

This paper presents a highly optimized, dual-stage pipeline for low-resource long-form Bangla speech processing, successfully tackling both semantic ASR chunking and complex speaker diarization.

For long-form ASR, we proposed a self-contained chunk-and-align pipeline that automatically derives utterance-level training segments from long-form recordings using per-word timestamp transfer and sequence matching---eliminating the need for external alignment tools or manual annotation. Fine-tuning the \texttt{tugstugi} Bengali Whisper checkpoint on these segments, combined with VAD-guided inference and hallucination post-processing, reduced WER from $67.5\%$ to $25.2\%$ on the public leaderboard.

For diarization, parameter-efficient fine-tuning of the segmentation module allowed the system to learn Bangla prosody, while migrating to the \texttt{community-1} architecture and utilizing \texttt{exclusive\_speaker\_diarization} natively resolved strict non-overlapping constraints without the need for destructive manual heuristics. By finally logically intersecting Pyannote's clustering with WhisperX's Silero VAD and enforcing a zero-second collar, we eliminated temporal boundary drift and ambient hallucinations. Ultimately, this computationally efficient framework dramatically outperformed baseline Bengali-Loop benchmarks, providing a robust, highly accurate blueprint for scaling speech technologies in low-resource environments.

\bibliographystyle{IEEEtran}
\bibliography{references}

@INPROCEEDINGS{10890105,
  author={Barański, Mateusz and Jasiński, Jan and Bartolewska, Julitta and Kacprzak, Stanisław and Witkowski, Marcin and Kowalczyk, Konrad},
  booktitle={ICASSP 2025 - 2025 IEEE International Conference on Acoustics, Speech and Signal Processing (ICASSP)},
  title={Investigation of Whisper ASR Hallucinations Induced by Non-Speech Audio},
  year={2025},
  volume={},
  number={},
  pages={1-5},
  doi={10.1109/ICASSP49660.2025.10890105},
}

@misc{tabib2026bengaliloopcommunitybenchmarkslongform,
      title={Bengali-Loop: Community Benchmarks for Long-Form Bangla ASR and Speaker Diarization}, 
      author={H. M. Shadman Tabib and Istiak Ahmmed Rifti and Abdullah Muhammed Amimul Ehsan and Somik Dasgupta and Md Zim Mim Siddiqee Sowdha and Abrar Jahin Sarker and Md. Rafiul Islam Nijamy and Tanvir Hossain and Mst. Metaly Khatun and Munzer Mahmood and Rakesh Debnath and Gourab Biswas and Asif Karim and Wahid Al Azad Navid and Masnoon Muztahid and Fuad Ahmed Udoy and Shahad Shahriar Rahman and Md. Tashdiqur Rahman Shifat and Most. Sonia Khatun and Mushfiqur Rahman and Md. Miraj Hasan and Anik Saha and Mohammad Ninad Mahmud Nobo and Soumik Bhattacharjee and Tusher Bhomik and Ahmmad Nur Swapnil and Shahriar Kabir},
      year={2026},
      eprint={2602.14291},
      archivePrefix={arXiv},
      primaryClass={cs.SD},
      url={https://arxiv.org/abs/2602.14291}, 
}

@article{bain2022whisperx,
  title={WhisperX: Time-Accurate Speech Transcription of Long-Form Audio},
  author={Bain, Max and Huh, Jaesung and Han, Tengda and Zisserman, Andrew},
  journal={INTERSPEECH 2023},
  year={2023}
}

\end{document}